\documentclass[twocolumn,showpacs,preprintnumbers,amsmath,amssymb,prb,superscriptaddress]{revtex4}

\usepackage{graphicx}
\usepackage{epsf}

\begin{document}

\title{Acoustoelectric effects in very high-mobility $p$-SiGe/Ge/SiGe heterostructure}

\author{I. L. Drichko}
\author{A. M. Diakonov}
\author{E.V. Lebedeva}
\author{I. Yu. Smirnov}
\email{ivan.smirnov & mail.ioffe.ru} \affiliation{A. F. Ioffe
Physico-Technical Institute of Russian Academy of Sciences, 194021
St.Petersburg, Russia}
\author{O. A. Mironov}
\affiliation{Warwick SEMINANO R\&D Centre, University of Warwick
Science Park, Coventry CV4 7EZ, UK}
\author{M. Kummer}
\author{H. von K\"{a}nel}
\affiliation{Laboratorium f$\ddot{u}$r Festk\"{o}rperphysik ETH
Z\"{u}rich, CH-8093 Z\"{u}rich Switzerland}
\affiliation{EpiSpeed SA, Technoparkstrasse 1, CH-8005
Z\"{u}rich Switzerland}
\date{\today}

\begin{abstract}
{Measurement results of the acoustoelectric effects (surface
acoustic waves (SAW) attenuation and velocity) in a high-mobility
$p$-SiGe/Ge/SiGe structure are presented. The structure was LEPECVD
grown with a two dimensional (2D) channel buried in the strained Ge
layer. The measurements were performed as a function of temperature
(1.5 - 4.2 K) and magnetic field (up to 8.4 T) at different SAW
intensities at frequencies 28 and 87 MHz. Shubnikov-de Haas-like
oscillations of both SAW attenuation and the velocity change have
been observed. Hole density and mobility, effective mass, quantum
and transport relaxation times, as well as the Dingle temperature
were measured with a method free of electric contacts. The effect of
heating of the 2D hole gas by the electric field of the SAW was
investigated. Energy relaxation time $\tau_{\varepsilon}$ and the
deformation potential constant determined.}
\end{abstract}


\maketitle

\section{Introduction}
\label{Introduction}

Two-dimensional semiconductor systems are usually a multilayer
structure with a 2-dimensional conducting layer in the depth of the
order of hundreds of angstrom from the surface. Determination the
key parameters of these structures, involves customarily the
measurement of transport properties such as resistance,
magnetoresistance, the Hall effect on a direct current. Measurements
of these effects require ohmic electrical contacts. The fabrication
of such contacts often requires the heating of the studied samples,
which can lead to irreversible changes of their properties. Besides
it requires a Hall bar to be configured to meet the geometrical
conditions for the correct measurement of effects. Hall bridge
manufacturing usually requires a high technology photolithography.
Moreover, the absence of contacts eliminates otherwise possible
carrier injection into the low-dimension interface from the contact
areas.

The present paper provides main heavy hole characteristics in a
high-mobility  $p$-SiGe/Ge/SiGe heterostructure. The results were
obtained with an acoustic method which is deprived of the
disadvantages mentioned above as it implies use of samples of
rectangular shape and the complete absence of electrical contacts.

\section{Experimental results}
\label{Experimental results}

\begin{figure}[ht]
\centerline{
\includegraphics[width=\columnwidth]{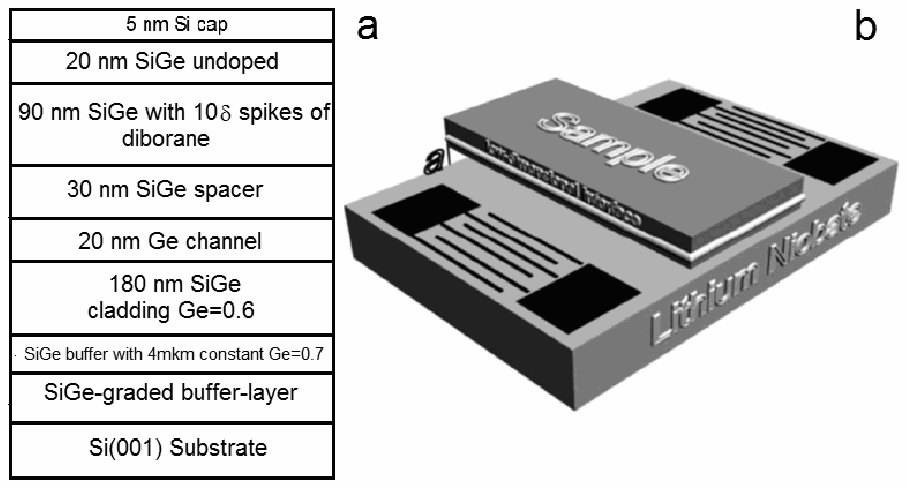}
} \caption{Sketches of sample (a) and the acoustic experiment setup (b).
\label{Sample}}
\end{figure}

\paragraph{Object}
The system under study was the sample $p$-SiGe/Ge/SiGe (K6016)
investigated earlier on a direct current in Ref.~\onlinecite{kanel}.
The sample structure is illustrated on Fig.~\ref{Sample}(a).

In the system under study the 2D-channel is in strained Ge. The
threefold degenerated  valence band of Ge is split due to a strain
into 3 subbands, top of which is occupied by heavy holes.

The entire structure was grown by low-energy plasma-enhanced
chemical vapor deposition (LEPECVD), by making use of the large
dynamic range of growth rates offered by that
technique~\onlinecite{kanel}. The buffer, graded at a rate of about
10$\% / \mu$m to a final Ge content of 70$\%$, and the 4 $\mu$m
thick constant composition layer were grown at a high rate of 5-10
nm/s while gradually lowering the substrate temperature $T_s$ from
720$^\text{o}$C to 450$^\text{o}$C. The active layer structure,
consisting of a pure 20 nm thick Ge layer sandwiched between
cladding layers with a Ge content of about 60$\%$ and a Si cap, was
grown at a low rate of about 0.3 nm/s at $T_s$ = 450$^\text{o}$C.
Modulation doping was achieved by introducing dilute diborane pulses
into the cladding layer above the channel after an undoped spacer
layer of about 30 nm.

\paragraph{Method} \label{Method}

The experimental setup is illustrated in Fig.~\ref{Sample}(b). It
includes a piezoelectric crystal (LiNbO$_3$), where a surface
acoustic wave (SAW) is excited at its surface by inter-digital
transducers. A SAW propagating along the surface of lithium niobate
is accompanied by high-frequency electric field. This electric field
penetrates into a 2DHG located in a semiconductor heterostructure
mounted on the surface. The field produces electrical currents
which, in turn, cause Joule losses. As a result, there are
electron-induced contributions both to the SAW attenuation and to
its velocity. These effects are governed by the complex
high-frequency conductivity, $\sigma^{AC}(\omega)$.

This "sandwich"-like experimental configuration allows to carry out
contactless acoustoelectric experiments on non-piezoelectric 2D
systems, for example, on SiGe/Ge/SiGe.

In the present experiment the acoustoelectric effects - attenuation
$\Gamma$ and the SAW velocity change $\Delta v$ - have been measured
at frequencies 28 and 87 MHz and a magnetic field up to 8.4 T in a
temperature range 1.6 - 4.2 K.

Fig.2 shows the magnetic field (B) dependences of the attenuation
$\Delta \Gamma (B) \equiv \Gamma(B)-\Gamma(0)$ and the SAW velocity
change $\Delta v(B)/v(0)  \equiv [v(B) - v(0)]/v(0)$ in the K6016
sample.

One can see that in magnetic field the absorption coefficient and
the velocity change both undergo Shubnikov-de Haas (SdH) type
oscillations, analogous to that in $\rho_{xx}$ on a direct current.
\begin{figure}[ht]
\centerline{
\includegraphics[width=7.5 cm]{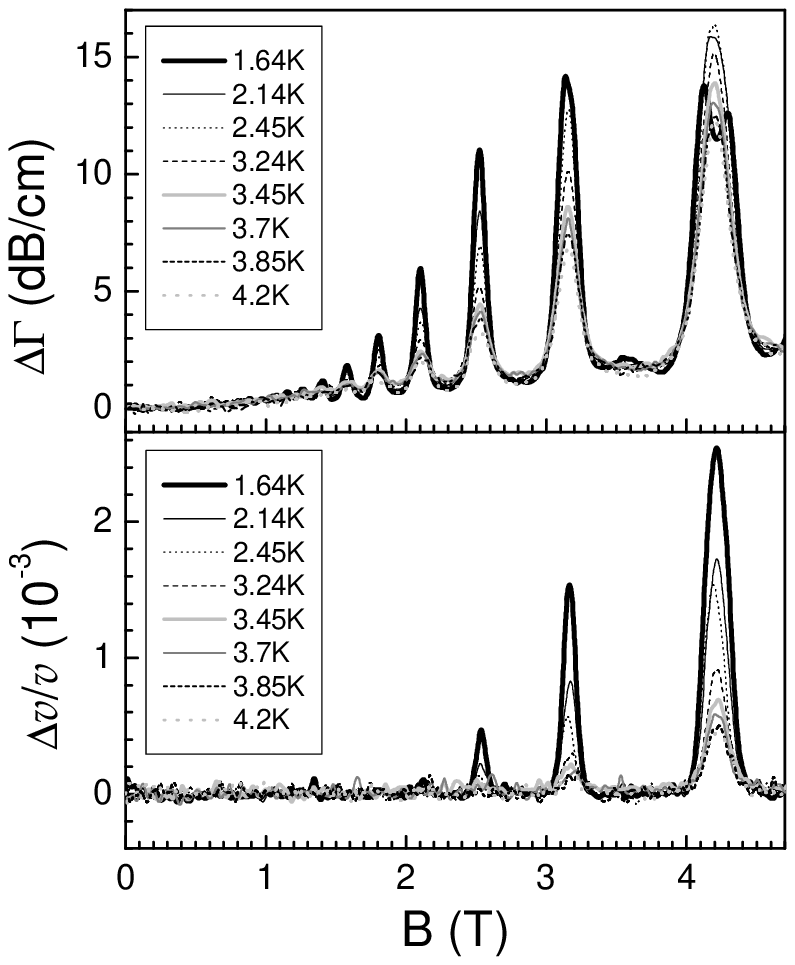}
} \caption{Magnetic field dependences of $\Delta \Gamma$ and
$\Delta
  v/v(0)$ for different temperatures. $f=28$ MHz.
 \label{fig:3}}
\end{figure}

From the measured values of $\Delta \Gamma$ and $\Delta v/v$ both
the real, $\sigma_1$ and imaginary, $\sigma_2$ parts of the
high-frequency (AC) conductivity $\sigma^{\text{AC}} (\omega) \equiv
\sigma_1-i\sigma_2$ could be obtained with the aid of
equations~(\ref{eq:G},\ref{eq:V}). Since $\Gamma(0) \ll \Gamma(B)$,
[\onlinecite{ildPRB}]:
\begin{eqnarray}
  \label{eq:G}
&&\Gamma=8.68\frac{K^2}{2}qA \times \nonumber \\
&&\times \frac{4\pi\sigma_1t(q)/\varepsilon_sv}
  {[1+4\pi\sigma_2t(q)/\varepsilon_sv]^2+[4\pi\sigma_1t(q)/\varepsilon_sv]^2}, \frac{\text{dB}}{\text{cm}}  \,   \\
&&A = 8b(q)(\varepsilon_1 +\varepsilon_0)
\varepsilon_0^2 \varepsilon_s
\exp [-2q(a+d)],  \,   \nonumber \\
\label{eq:V}
&&\frac{\Delta v}{v}=\frac{K^2}{2}A \times \nonumber \\
&&\times \frac{1+4\pi\sigma_2t(q)/\varepsilon_sv}
  {[1+4\pi\sigma_2t(q)/\varepsilon_sv]^2+[4\pi\sigma_1t(q)/\varepsilon_sv]^2},
\end{eqnarray}
where $K^2$ is the electro-mechanic coupling constant for lithium
niobate (Y-cut),
$q$ and $v$ are the SAW wave vector and velocity in LiNbO$_3$, respectively. $a$
is the gap between the piezoelectric plate and the sample, $d$ distance between the sample
surface and the 2DHG layer; $\varepsilon_1$,
$\varepsilon_0$ and $\varepsilon_s$ are the dielectric constants of
LiNbO$_3$, of
 vacuum, and of the semiconductor, respectively; $b$ and $t$ are complex
functions of $a$, $d$, $\varepsilon_1$, $\varepsilon_0$ and
$\varepsilon_s$. In order to obtain $\sigma_1$ and $\sigma_2$ from
equations~(\ref{eq:G},\ref{eq:V}) one should know the values of $a$
and $d$. $d\approx$1.45$\times 10^{-5}$ cm is determined in the
technological process of the sample production, and $a$ is
determined by fitting the experimental data at those magnetic fields
where the conductance is metallic and essentially frequency
independent. The value of $a$ depends on the assembling quality of
the sample-LiNbO$_3$ sandwich, and was in this experiment $a\approx
3.1 \times 10^{-5}$ cm. AC conductivity components $\sigma_1$ and
$\sigma_2$ as a function of magnetic field were measured at
different temperatures.

\begin{figure}[t]
\centerline{
\includegraphics[width=8cm,clip=]{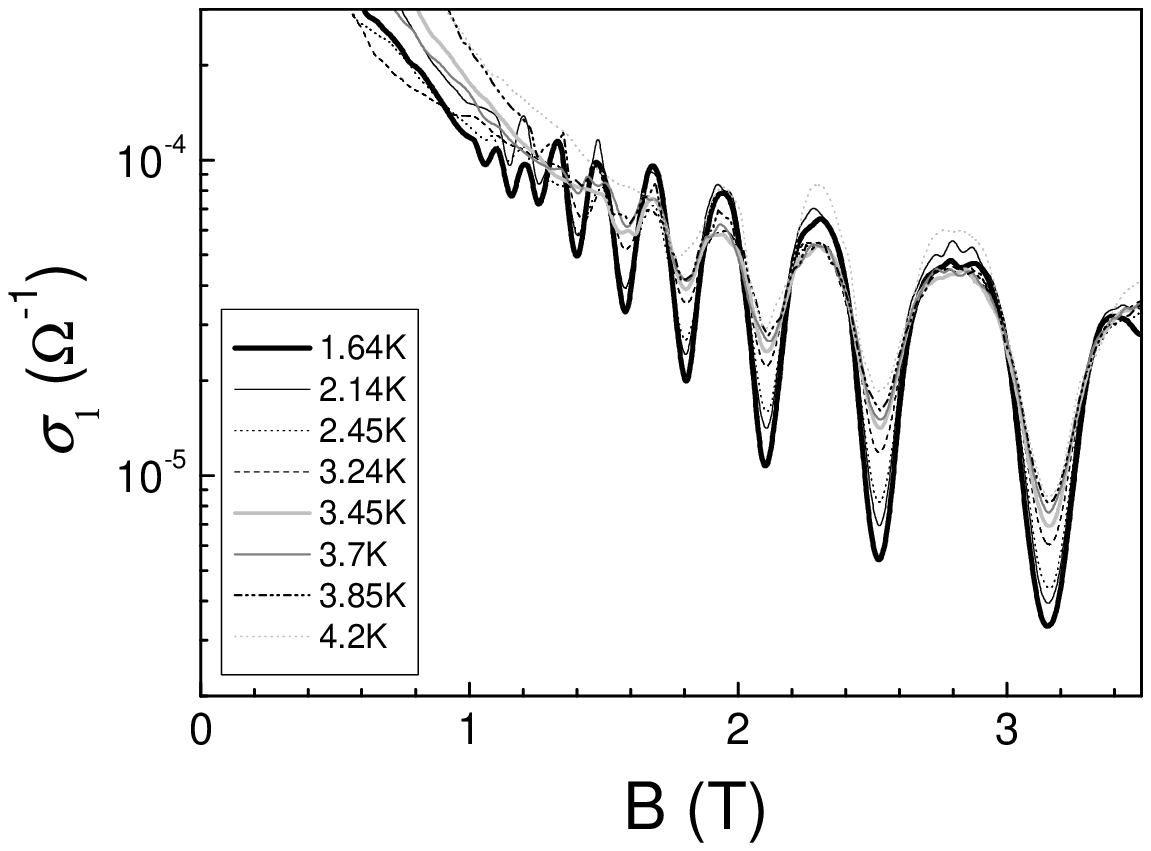}} \caption{Dependence of the real component of the
AC conductivity at 28 MHz on a magnetic field at different temperatures.}
\label{S1}
\end{figure}

Fig.~\ref{S1} illustrates the magnetic field dependence of the
real $\sigma_{1}$ component of the
complex high-frequency conductivity, obtained from SAW
measurements at different temperatures using Eqs.~(\ref{eq:G},\ref{eq:V}).

\paragraph{Determination of parameters} \label{Determination of
parameters}\

1) The density of holes, obtained from the SdH oscillations minima
appeared to be: $p=(6.1 \pm 0.1) \times 10^{11}$ cm$^{-2}$

2) Mobility temperature dependence at $B$=0.

According to Ando's theory [\onlinecite{ando}], $\sigma_{xx}$ in a
magnetic field can be expressed in a form:
\begin{equation}
  \label{eq:01x}
  \sigma_{xx}=\sigma_{xx}^*+\sigma_{xx}^{osc},
\end{equation}
where $\sigma_{xx}^*=\frac{\sigma_{0}}{1+\omega_c^2\tau_0^2}$ is the
classical Drude conductivity, $\sigma_{0}$ is the zero magnetic
field conductivity, $\omega_c=eB/m^*c$ is the cyclotron frequency,
$\tau_0$ is the transport relaxation time; $\omega_c \tau_0 =\mu
B/c$, where $\mu$ is the mobility, $c$=3$\times 10^{10}$cm/s.

All acoustic method measurements are relative, thus it is impossible
to obtain from them directly the absolute values of either
conductivity, or mobility at $B$=0. However, in case of high
mobility samples, where there are many SdH oscillations in the
relatively low magnetic field region (no localization) these values
could be determined. In this case one could plot $\sigma_{xx}^*$ -
mean value between the oscillations envelope, as a function of
$1/B^2$. The mobility at $B$=0 is determined from the slope of this
line.
\begin{equation}
  \label{eq:02}
  \sigma_{xx}^*=\frac{\sigma_{0}}
  {1+\omega_c^2\tau_0^2}=\frac{epc^2}{\mu B^2}, \text{ provided } (\omega_c^2\tau_0^2)\gg 1.
\end{equation}

\begin{figure}[h]
\centerline{
\includegraphics[width=7cm,clip=]{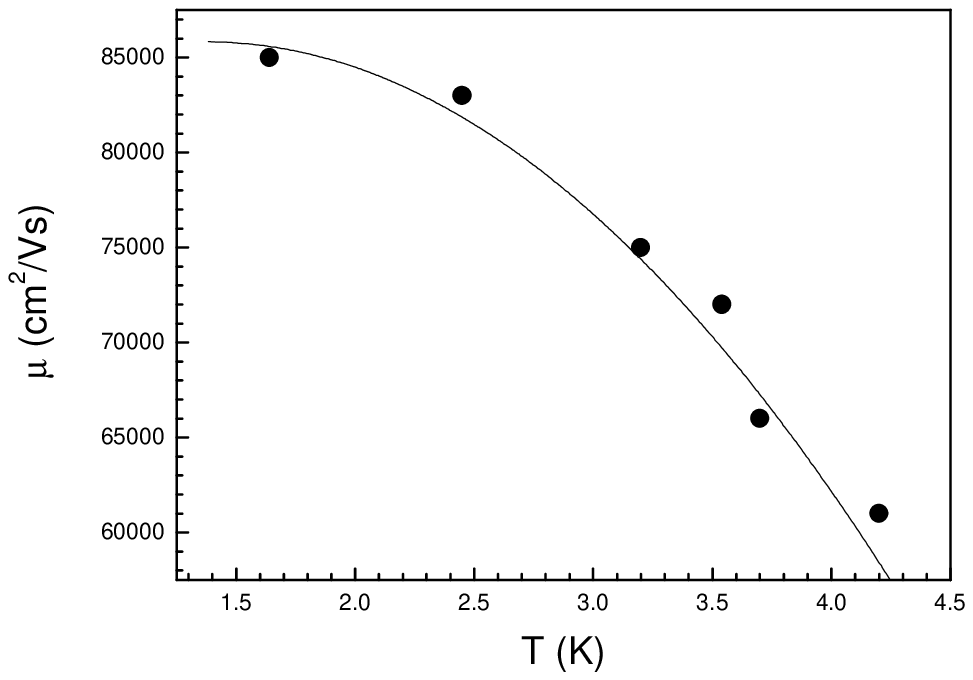}}
\caption{Dependence of mobility, $\mu$ at $B$=0 on a temperature. (The curve plotted as judged by eye).}
\label{mobil}
\end{figure}

Fig.~\ref{mobil} presents the mobility value, obtained with the
method described as a function of temperature.

The $\sigma_{0}$ value can be determined with the aid  of the
density $p$ and mobility, $\mu$ obtained in our measurements. For
$T$=1.7 K and $B$=0 $\sigma_{0}=ep\mu$=0.008 Ohm$^{-1}$ which value
by only 5$\%$ differs from $\sigma_{0}$=0.0076 Ohm$^{-1}$  measured
in Ref.~\onlinecite{kanel}.

3) Determination of the effective mass and of the ratio
$\tau_0/\tau_q$, where $\tau_q$ is the quantum relaxation time.

\begin{eqnarray}
  \label{eq:03}
  &&\sigma_{xx}^{osc}\sim \sigma_{xx}^{*} D(X_T)\exp(-\pi/\omega_c \tau_q)
  \cos(2\pi E_F/\hbar\omega_c - \pi) \, ,  \nonumber \\
  &&D(X_T)=X(T)/\sinh X(T), X(T)=2\pi^2 k_B T/\hbar \omega_c,
\end{eqnarray}
where $E_F$ is the Fermi energy.

The effective mass is conventionally determined from the temperature
dependences of the oscillation amplitude at different magnetic
fields  in the 1.5 - 2.1 T range. The $\tau_0/\tau_q$ ratio obtains
from the oscillation amplitude dependence on $1/\omega_c \tau_0$ at
different temperatures. As a result, one has $m^*=(0.11 \pm
0.01)m_0$, $\tau_0/\tau_q$=4.9$\pm$0.2. According to
Ref.~\onlinecite{gold}, this corresponds to the holes scattering at
the charged impurity centers with the screening  taken into account.
The values of $p$, $m^*$, and $\tau_0/\tau_q$ thus obtained are very
close to those of Ref.~\onlinecite{kanel}.

Absolute values at $T$-1.7 K are: the transport relaxation time
$\tau_0=(4.8 \pm 0.3)\times 10^{-12}$ s, the quantum relaxation time
$\tau_q=(1.0 \pm 0.1)\times 10^{-12}$ s and the Dingle temperature
$T_D=\hbar/2\pi\tau_q$=1.24 K.

4) Energy relaxation time $\tau_\varepsilon$.

Additional measurements of the acousto-electronic effects dependence
on the SAW power at 1.6 K in a magnetic field up to 2.5 T were
performed for the determination of the relaxation time of energy .

\begin{figure}[h]
\centerline{
\includegraphics[width=7.5cm,clip=]{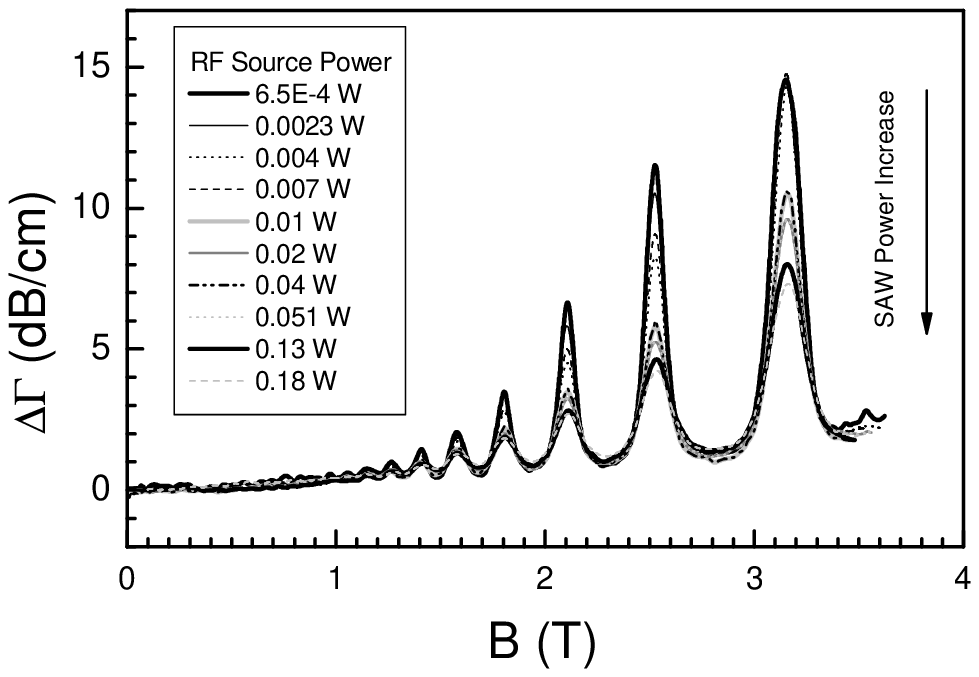}}
\caption{Dependences of $\Delta\Gamma$ on magnetic field B at different
rf-source powers, $T$=1.5 K, $f$ =30 MHz..}
\label{Gpower}
\end{figure}

Fig.~\ref{Gpower} depicts the attenuation, $\Gamma$, dependence on
magnetic field for diverse SAW powers at the source generator
output.

As it is easily seen in Fig.~\ref{Gpower}, $\Gamma$ depends on the
SAW power introduced into a sample. Since all holes parameters
obtained above were measured  in their delocalized state, it seems
natural to suppose that in the same magnetic field region the cause
of $\Gamma$ dependence on SAW power is the 2DHG heating in strong
electric field of the SAW. Following Ref.~\onlinecite{HeatSAW}, we
shall use the concept of the hole gas temperature $T_e$ and
determine it comparing  two dependences: $\Gamma(P)$ and
$\Gamma(T_0)$, where $P$ is the RF source power, $T_0$ is the
lattice temperature. This comparison enables one to establish
correspondence between the hole gas temperature and the RF source
output power. The energy loss rate, $Q$, dependence on the hole gas
temperature was plotted for the determination of the relaxation
mechanism.

According to Ref.~\onlinecite{HeatSAW}, the energy losses rate
\begin{equation}
  \label{eq:04}
 Q=\sigma_{xx}E^2=4W \Gamma,
\end{equation}
where  $E$ is the electric field produced by the SAW and penetrating
the channel with the 2D holes, $\Gamma$ is the attenuation and $W$
is the SAW power at the sample entrance. $W$ is determined as the
output power of the RF source generator minus the loss associated
with the electro-acoustic conversion and transmission line.
\begin{figure}[h]
\centerline{
\includegraphics[width=7cm,clip=]{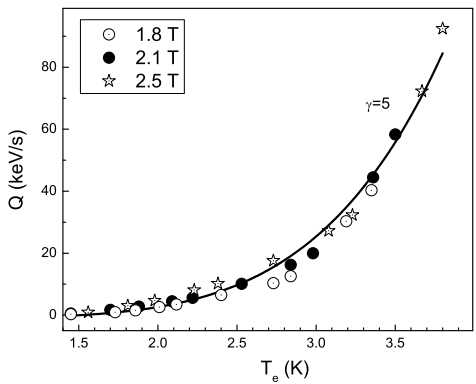}}
\caption{The energy losses rate per hole measured in acoustics $Q$
plotted versus $T_e$. Line is the fitting curve with $Q=A_\gamma (T_e^{\gamma}-T_0^{\gamma})$,
$\gamma$=5.}
\label{QTe}
\end{figure}

$Q=A_5 (T_e^{5}-T_0^{5})$ relation seems the most suitable for
experimental data interpretation. Here $A_5$ could be determined
from the slope of the $Q(T_e^{5}-T_0^{5})$. The dependence of this
kind evidences that the energy relaxation is determined by the
unscreened acoustic phonon deformation potential scattering. At the
limit of weak heating, when $\Delta T < T_0$, the energetic
relaxation time could be estimated, following
Ref.~\onlinecite{HeatSAW}:
\begin{equation}
  \label{eq:05}
 \tau_{\varepsilon}=\frac{\pi^2 k_B^2}{3\gamma A_{\gamma} E_FT_0^{\gamma-2}}.
\end{equation}
In our case, when $\gamma$=5, one gets:
$\tau_{\varepsilon}$=$(6.3\pm 0.1)\times$10$^{-10}$ s.

Regarding the energy relaxation as a deformation potential
scattering with a weak screening, one can determine the deformation
potential value. Indeed, according to Ref.~\onlinecite{Karpus}, the
energetic loss value could then be evaluated with the aid of a
formula:
\begin{eqnarray}
  \label{eq:07}
  &&Q_{DA}=b_2 Q_2(\frac{k_B T_0}{2m^* s^2})^2 (\frac{k_B T_0}{\hbar k_F s})^3
  (\frac{T_e^5}{T_0^5}-1)\, ,  \nonumber \\
  &&Q_2 \equiv \frac{2m^*s^2}{l_o / s}, b_2=12 \zeta(5)\, ,
  l_0 \equiv \frac{\pi \hbar^4 \rho}{2m^{*3} E_{DA}^2},
\end{eqnarray}
where $s$ and $\rho$ are the longitudinal sound velocity and the
density of Ge, $k_F$ is the wavevector of the electron with an
energy of Fermi ($E_F$), $\zeta$(x) is the Rihman function, $E_{DA}$
is the deformation potential. $E_{DA}$=(12$\pm$1) eV.

Deformation potential of unstressed Ge, determined in
Ref.~\onlinecite{deform}, $E_{DA}$=(16.2$\pm$0.4) eV.

\begin{figure}[h]
\centerline{
\includegraphics[width=7.5cm,clip=]{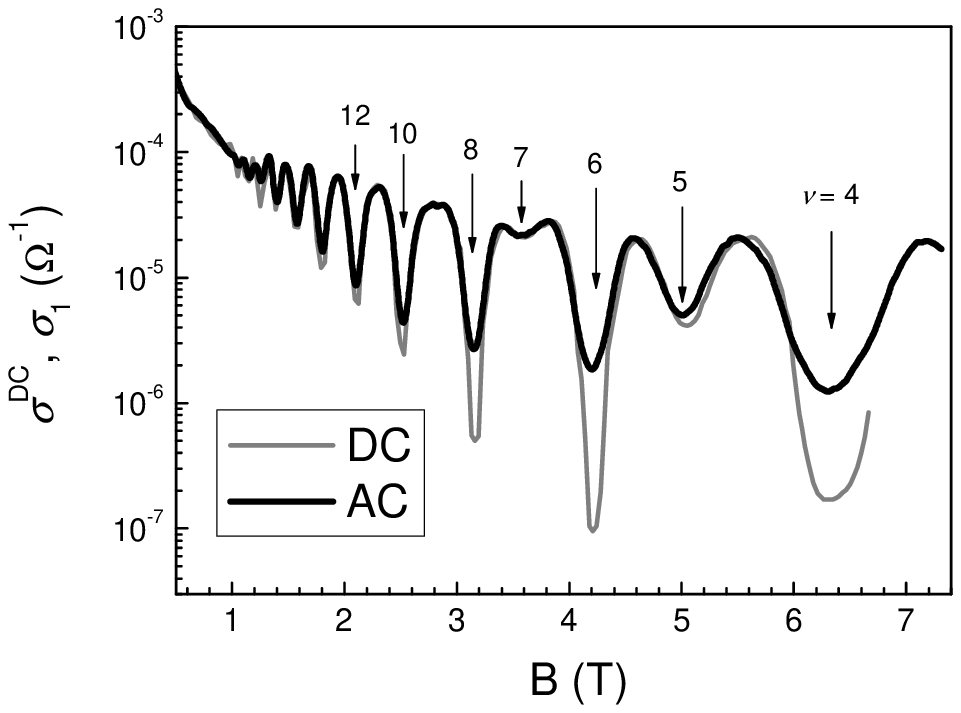}}
\caption{Magnetic field dependence of $\sigma^{DC}_{xx}$ (grey line), and the real part $\sigma_1$ of the AC conductivity, $f$=28 MHz, $T$ = 1.7 K. }
\label{AcDc}
\end{figure}

As a conclusion, in Fig.~\ref{AcDc} presented are the magnetic field
dependence of both DC conductivity from [\onlinecite{kanel}], and
that of the real part, $\sigma_1$, of the AC conductivity at $T$ =
1.7 K.

$\sigma^{DC}_{xx}$ was obtained from Fig.3 of Ref.~\onlinecite{kanel} as $\sigma^{DC}_{xx}=\frac{\rho_{xx}}{\rho_{xx}^2+\rho_{xy}^2}$.

As one can conclude observing the figure, the curves at $T$ = 1.7 K
practically coincide up to $B$ =2.1 T, and then $\sigma_1$ exceeds
$\sigma^{DC}_{xx}$ and the difference between these values grows
with the growth of the magnetic field. In relatively week magnetic
field the carriers are delocalized $\sigma^{DC}_{xx}$=$\sigma_1$ and
$\sigma_2$=0 [\onlinecite{loc}], the conductivity mechanisms for DC
and AC are the same. With the increase of the magnetic field  the
holes in the SdH oscillation minima became localized, and the
conductivity turns to be hopping. DC hopping differs from hopping in
AC. In the DC case hopping is accomplished  to the nearest neighbour
with a constant activation energy, than changing to the variable
length hopping (see [\onlinecite{Greenbook}]). In the alternating
electric field the hopping conductivity is treated with the "two
site model", in this case always $\sigma^{DC}_{xx} \ll \sigma_1$
[\onlinecite{ildPRB,Pollak,Efros}]. As soon as the 2DHG parameters
should be determined  when the carriers are not localized, the curve
of Fig.~\ref{AcDc} informs that it could be done at magnetic fields,
not exceeding 2.1 T.

\section{Conclusion}
\label{Conclusion}

In conclusion we draw here a comparative table with data obtained
via conventional dc measurements [\onlinecite{kanel}] and using contactless SAW technique.

\begin{table}[h]\begin{center}
\caption{Comparative table with data obtained from dc measurements
[\onlinecite{kanel}] and using contactless SAW technique.}
\label{DCAC}

\begin{tabular}{p{1.5cm} @{\hspace{.01cm}} p{3.3cm}@{\hspace{.01cm}} p{3.5cm}}

\hline
   & DC & AC \\ \hline
  $p$(cm$^{-2}$)&$(6.075\pm0.005)\times10^{11}$&$(6.1\pm0.1)\times 10^{11}$\\
  $m^*$ & $(0.104\pm0.001)m_0$ &$(0.11\pm0.01)m_0$ \\
  $\mu_{1.7K}$\\(cm$^2$/Vs)&$7.8\times10^4$&$(8.5\pm0.6)\times10^4$\\
  $\tau_0/\tau_q$ & 7.5 & 4.9$\pm$0.2 \\
  $\tau_{\varepsilon}$ (s)& - &(6.3$\pm$0.1)$\times$10$^{-10}$ \\
  $E_{DA}$ (eV)& - &12$\pm$1\\
\hline
\end{tabular}
\end{center}
\end{table}

\acknowledgments This work was supported  by grants of RFBR
08-02-00852; the Presidium of the Russian Academy of Science, the
Program of Branch of Physical Sciences of RAS "Spintronika".

\end{document}